\documentclass[epsfig,aps,pra,preprint]{revtex4}
\usepackage{graphicx}
\usepackage{amsmath}

\begin{document}

\draft \preprint{}

\title{A Finite Temperature Treatment of Ultracold Atoms in a 1-D Optical
Lattice}

\author{B. G. Wild, P. B. Blakie and D. A. W. Hutchinson}

\address{Department of Physics,
University of Otago,
Dunedin, New Zealand}

\begin{abstract}
We consider the effects of temperature upon the superfluid phase of ultracold, weakly
interacting bosons in a one dimensional optical lattice. We use a finite temperature
treatment of the Bose-Hubbard model based upon the Hartree-Fock-Bogoliubov formalism,
considering both a translationally invariant lattice and one with additional harmonic
confinement. In both cases we observe an upward shift in the critical temperature for
Bose condensation. For the case with additional harmonic confinement, this is in contrast
with results for the uniform gas.
\end{abstract}

\maketitle

\section{Introduction}

Ultracold atoms confined within optical lattice potentials are of great current interest,
both theoretically\cite{key-1,key-2,key-3,key-4,key-5} and experimentally\cite{key-6,key-7}. Of particular interest are the
connections to solid state regimes where optical lattices can be manufactured with
extraordinary control so as to simulate more complicated, less perfect and, often, less
manipulable systems, initially of interest in a condensed matter sphere
(eg. Heisenberg or Ising Hamiltonians )\cite{key-8,key-9,key-10}.  Ultracold atoms in 
optical lattices have also been proposed as candidates for
quantum information processing\cite{key-11}, which has applications
in quantum cryptography and quantum computing. Various methods of
loading Bose-Einstein concensates into optical lattices have been
proposed\cite{key-12,key-13,key-14,key-15}, and this is now routinely performed\cite{key-14},
as is the manipulation and control of the atoms in such a lattice\cite{key-16}.

The focus of this paper is on the microscopic treatment of Bose-Einstein
condensates in a one dimensional optical lattice. 
We extend previous treatments at zero temperature\cite{key-2,key-3} to finite 
temperature using the 
Hartree-Fock-Bogoliubov mean-field treatment as applied to a discrete
Bose-Hubbard model yielding modified Gross-Pitaevskii
and Bogoliubov-de Gennes equations. This set of
equations is solved for both the case of a translationally invariant
lattice (no external trapping potential) and an inhomogeneous lattice
(optical lattice in an external harmonic trapping potential). The use of the mean field
based treatment means we are only considering the superfluid phase for the atoms in the
optical lattice. The model is not valid in the Mott Insulator regime. We use the model to
estimate the superfluid to normal phase
transition temperature in each case. Thus one is able to obtain a phase diagram for
the superfluid and normal gas states in the low effective interaction strength limit.

\section{Formalism}

We consider a one-dimensional optical lattice with $I$ lattice
sites. We begin from the Bose-Hubbard
Hamiltonian for atoms in a one-dimensional
optical lattice \cite{key-1,key-2,key-3}
\begin{equation}
\hat{H}=\sum_{i=1}^{I}\hat{n}_{i}\epsilon_{i}-J\sum_{i=1}^{I}\left(\hat{a}_{i+1}^{\dag}\hat{a}_{i}+\hat{a}_{i}^{\dag}\hat{a}_{i+1}\right)+\frac{V}{2}\sum_{i=1}^{I}\hat{n}_{i}\left(\hat{n}_{i}-1\right),\label{eq1}
\end{equation}
 where $J$ represents the coupling strength between adjacent lattice
sites, $V$ is the interaction potential acting between atoms on the
same site, and $\hat{a}_{i}$ is the Bose field operator for the $i^{th}$
lattice site, and $\hat{n}_{i}=\hat{a}_{i}^{\dag}\hat{a}_{i}$ . $\epsilon_{i}$
is the energy on each lattice site $i$ due to the trapping potential. The
usual commutation relations apply for the Bose field operator $\hat{a}_{i}$.
Assuming a macroscopic occupation of the ground state, we express
the Bose annihilation operator for each lattice site $i$ in terms
of a complex mean field part $z_{i}$ and a fluctuation operator part $\hat{\delta}_{i}$
\cite{key-17}, $\hat{a}_{i}=\left(z_{i}+\hat{\delta}_{i}\right)\exp\left(\frac{-i\mu t}{\hbar}\right)$, where $\mu$ is the eigenvalue for the generalised Gross-Pitaevskii equation to be discussed below, and take the self-consistent mean-field approximation such that 
$\hat{\delta}_{i}^{\dag}\hat{\delta}_{i}\hat{\delta}_{i}=2\left\langle \hat{\delta}_{i}^{\dag}\hat{\delta}_{i}
\right\rangle \hat{\delta}_{i}+\left\langle \hat{\delta}_{i}\hat{\delta}_{i}\right\rangle \hat{\delta}_{i}^{\dag}$,
etc. The condensate density is given by $n_{c_{i}}=\left|z_{i}\right|^{2}$
and $\tilde{n}_{i}=\left\langle \hat{\delta}_{i}^{\dag}\hat{\delta}_{i}\right\rangle $ is
the excited population density. The anomalous density is neglected\cite{key-3,key-17,key-18,key-19}.

The resulting Hamiltonian can now be diagonalised using the Bogoliubov (canonical) transformation\begin{equation}
\hat{\delta}_{i}=\sum_{q}\left[u_{i}^{q}\hat{\alpha}_{q}e^{-i\omega_{q}t}-v_{i}^{q^{*}}\hat{\alpha}_{q}^{\dag}e^{i\omega_{q}t}\right]\label{eq10}\end{equation}
and

\begin{equation}
\hat{\delta}_{i}^{\dag}=\sum_{q}\left[u_{i}^{q^{*}}\hat{\alpha}_{q}^{\dag}e^{i\omega_{q}t}-v_{i}^{q}\hat{\alpha}_{q}e^{-i\omega_{q}t}\right],\label{eq11}\end{equation}
 where $u_{i}^{q}$ and $v_{i}^{q}$ are the quasiparticle amplitudes,
$\omega_{q}$ are the quasiparticle excitation frequencies, and $\hat{\alpha}_{q}^{\dag}$
 ($\hat{\alpha}_{q}$) is the quasiparticle creation (annihilation) operator. This yields
\cite{key-3} a set
of coupled equations comprising of a modified
Gross-Pitaevskii equation
\begin{equation}
\mu^{\prime}z_{i}=\epsilon_{i}^{\prime}z_{i}-\left(z_{i+1}+z_{i-1}\right)+V_{\rm{eff}}\left(n_{c_{i}}z_{i}+2\tilde{n}_{i}z_{i}+\tilde{m}_{i}z_{i}\right)\label{eq12}
\end{equation}
and the Bogoliubov-de Gennes equations
\begin{equation}
\begin{array}{c}
\hbar\omega_{q}^{\prime}u_{i}^{q}+c_{q}^{\prime}z_{i}=\left[2V_{\rm{eff}}\left(n_{c_{i}}+\tilde{n}_{i}\right)-\mu^{\prime}+\omega_{i}^{\prime}\right]u_{i}^{q}-\left[u_{i+1}^{q}+u_{i-1}^{q}\right]-V_{\rm{eff}}z_{i}^{2}v_{i}^{q}\\
-\hbar\omega_{q}^{\prime}v_{i}^{q}-c_{q}^{\prime}z_{i}=\left[2V_{\rm{eff}}\left(n_{c_{i}}+\tilde{n}_{i}\right)-\mu^{\prime}+\epsilon_{i}^{\prime}\right]v_{i}^{q}-\left[v_{i+1}^{q}+v_{i-1}^{q}\right]-V_{\rm{eff}}z_{i}^{2}u_{i}^{q}\end{array},\label{eq13}
\end{equation}
 with
 \begin{equation}
\tilde{n}_{i}=\left\langle \hat{\delta}_{i}^{\dag}\hat{\delta}_{i}\right\rangle =\sum_{q}\left[\left|v_{i}^{q}\right|^{2}+\left(\left|u_{i}^{q}\right|^{2}+\left|v_{i}^{q}\right|^{2}\right)N_{BE}\left(\hbar\omega_{q}^{\prime}\right)\right]\label{eq15}
\end{equation}
 where
 \begin{equation}
N_{BE}\left(\hbar\omega_{q}^{\prime}\right)=\frac{1}{Z^{-1}e^{\beta\hbar\omega_{q}^{\prime}}-1}\label{eq16}
\end{equation}
is the usual Bose distribution, the primed quantities are measured in units of $J$ (which depends on the depth of the optical lattice), and where we have assumed without loss of
generality the condensate amplitudes $z_{i}$ to be real. $\mu$ is the energy eigenvalue
for the modified Gross-Pitaevskii equation (and approximates the chemical
potential closely for values of temperature well below the transition
temperature) and $\hbar\omega_{q}$ is the energy eigenvalue of the Bogoliubov-de Gennes equations. The $c^{q}$'s
are necessary in order to ensure the orthogonality of the condensate
with the excited states\cite{key-3,key-18}. $Z$ is a fugacity term resulting from the difference between the
true chemical potential $\mu_{T}$, and the chemical potential as
estimated using the eigenvalue $\mu$ corresponding to the ground
state of the modified Gross-Pitaevskii equation.
Thus $Z=\exp\left(\beta\left(\mu_{T}-\mu\right)\right)$. 

In the case of a homogeneous gas with a large
number of particles in the ground state (ie. $n_{c}\gg1$), the fugacity
may be approximated by\cite{key-18}\begin{equation}
Z=1+1/n_{c}\label{eq17}\end{equation}
 In the lattice, this is not always a good approximation, since the condition $n_{c}\gg1$
does not always hold, and we will take $Z=1$. This will lead to increased
values for the excited atom population, and will therefore result
in under-estimated values of the transition temperature.

Superfluid flow of the condensate occurs when there is a phase gradient.
A phase gradient of the condensate modifies the hopping term of the
Hamiltonian by the introduction of Peierls phase factors. The resulting
energy shift may be estimated using second order perturbation theory.
Since this energy shift is due entirely to the kinetic energy associated
with the superfluid flow, and hence due to the superfluid fraction,
it follows that the superfluid fraction $f_{s}$ may be calculated
in terms of this energy shift\cite{key-3}
\begin{equation}
f_{s}=\frac{1}{N}\frac{E_{\phi}-E_{0}}{J\left(\Delta\phi\right)^{2}}=-\frac{1}{2NJ}\left\langle \psi_{0}\right|\hat{T}\left|\psi_{0}\right\rangle -\frac{1}{NJ}\sum_{\nu\neq0}\frac{\left|\left\langle \psi_{\nu}\right|\hat{J}\left|\psi_{\nu}\right\rangle \right|^{2}}{E_{\nu}-E_{0}}
\end{equation}
where $\hat{J}=iJ\sum_{i=1}^{I}\left(\hat{a}_{i+1}^{\dag}\hat{a}_{i}-\hat{a}_{i}^{\dag}\hat{a}_{i+1}\right)$
and $\hat{T}=J\sum_{i=1}^{I}\left(\hat{a}_{i+1}^{\dag}\hat{a}_{i}+\hat{a}_{i}^{\dag}\hat{a}_{i+1}\right)$.
The superfluid fraction can then be expressed
in terms of the condensate and quasiparticle amplitudes
\begin{equation}
f_{s}=f_{s}^{(1)}-f_{s}^{(2)}\label{eq20}
\end{equation}
where
\begin{equation}
f_{s}^{(1)}=\frac{1}{2N}\sum_{i=1}^{I}\left[\left(z_{i+1}z_{i}^{*}+z_{i+1}^{*}z_{i}\right)+\sum_{q}\left(v_{i}^{q}v_{i+1}^{q^{*}}+v_{i}^{q^{*}}v_{i+1}^{q}\right)\right]\label{eq21}
\end{equation}
and
\begin{equation}
f_{s}^{(2)}=\frac{J}{N}\sum_{q,q^{\prime}}\left[\frac{\left|\sum_{i}\left(u_{i+1}^{q}v_{i}^{q^{\prime}}-u_{i}^{q}v_{i+1}^{q^{\prime}}\right)\right|^{2}}{\hbar\omega_{q}+\hbar\omega_{q}^{\prime}}+\delta_{qq^{\prime}}\frac{\left|\sum_{i}\left(u_{i+1}^{q}v_{i}^{q}-u_{i}^{q}v_{i+1}^{q}\right)\right|^{2}}{2\hbar\omega_{q}}\right].\label{eq22}
\end{equation}

\subsection{Translationally Invariant Lattice\label{sub:3.1}}

In the case of a translationally invariant lattice, periodic boundary
conditions apply, and the quasi-particle amplitudes are given by
\begin{equation}
\begin{array}{ccccc}
u_{j}^{q}=\frac{u^{q}e^{i(qja)}}{\sqrt{I}} & , & v_{j}^{q}=\frac{v^{q}e^{i(qja)}}{\sqrt{I}} & , & 1\leq j\leq I-1\end{array}\label{eq32}
\end{equation}
 Furthermore the condensate amplitudes are equal for each site $j$
and, since there is no trapping potential, $\epsilon_{i}=\epsilon=0$. Solving for the 
quasiparticle amplitudes we obtain 
\begin{equation}
\left|u^{q}\right|^{2}=\frac{Vz^{2}+4J\sin^{2}\left(\frac{qa}{2}\right)+\hbar\omega_{q}}{2\hbar\omega_{q}}\label{eq36}\end{equation}
\begin{equation}
\left|v^{q}\right|^{2}=\frac{Vz^{2}+4J\sin^{2}\left(\frac{qa}{2}\right)-\hbar\omega_{q}}{2\hbar\omega_{q}},\label{eq37}
\end{equation}
 and
\begin{equation}
\hbar\omega_{q}=\sqrt{4J\sin^{2}\left(\frac{qa}{2}\right)\left[2n_{c}V+4J\sin^{2}\left(\frac{qa}{2}\right)\right]}.\label{eq38}\end{equation}
 Thus from equations (\ref{eq36}), (\ref{eq37}) and (\ref{eq38}),
for the translationally invariant lattice, the superfluid fraction
is given by\begin{equation}
f_{s}=\frac{1}{N}\left(I\left|z\right|^{2}+\sum_{j=1}^{I-1}\left|v^{q}\right|^{2}\cos\left(\frac{2\pi j}{I}\right)\right),\label{eq43}\end{equation}
 since $q=2\pi j/Ia$ for $1\leq j\leq I-1$.
The condensate fraction for a given lattice site is given by $f_{c}=n_{c}/n_{0}$
where $n_{0}=N/I$ is the number of atoms per site. 

To calculate the condensate and superfluid fraction, we first determine the condensate amplitude. Then
defining\begin{equation}
g_{j(n)}=\left|z\right|_{(n-1)}^{2}V_{\rm{eff}}+4\sin^{2}\left(\frac{\pi j}{I}\right)\label{eq49}\end{equation}
\begin{equation}
e_{j(n)}=2\sin\left(\frac{\pi j}{I}\right)\sqrt{2\left|z\right|_{(n-1)}^{2}V_{\rm{eff}}+4\sin^{2}\left(\frac{\pi j}{I}\right)}\label{eq50}\end{equation}
 and\begin{equation}
N_{BE_{j(n)}}=\frac{1}{\exp\left(\beta^{\prime}e_{j_{(n-1)}}\right)-1},\label{eq51}\end{equation}
 where $\beta^{\prime}=J/k_{B}T$,
$k_{B}$ is Boltzman's constant and where the subscript
$(n)$ refers to the variable in question at the $n^{\mathrm{th}}$
iteration, one can solve for $\left|z\right|^{2}$ using an iterative
scheme\begin{equation}
\left|z\right|_{(n)}^{2}=\frac{1}{I}\left(N-\sum_{j=1}^{I-1}\left[\frac{g_{j(n)}-e_{j(n)}}{2e_{j(n)}}+\frac{g_{j(n)}}{e_{j(n)}}N_{BE_{j(n)}}\right]\right).\label{eq52}\end{equation}
 As an initial guess, the value $\left|z\right|_{(n)}^{2}=\frac{N}{I}$
is used, and the calculations (\ref{eq49}), (\ref{eq50}), (\ref{eq51})
and (\ref{eq52}) repeated until convergence is attained (ie. $\left|\left|z\right|_{(n)}^{2}-\left|z\right|_{(n-1)}^{2}\right|<$
Error Tolerance), or the maximum number of iterations is exceeded
(divergent solution)). 

One first calculates $g_{j(n)}$, $e_{j(n)}$ and $N_{BE_{j(n)}}$using
equations (\ref{eq49}), (\ref{eq50}) and (\ref{eq51}). The quasiparticle
amplitudes (given by equations (\ref{eq36}) and (\ref{eq37})) may
be calculated using the equations\begin{equation}
\left|u^{(j)}\right|_{(n)}^{2}=\frac{g_{j(n)}+e_{j(n)}}{2e_{j(n)}}\label{eq53}\end{equation}
 and\begin{equation}
\left|v^{(j)}\right|_{(n)}^{2}=\frac{g_{j(n)}-e_{j(n)}}{2e_{j(n)}}\label{eq54}\end{equation}
 The condensate and superfluid fractions are then readily determined.

\subsection{Inhomogeneous Lattice\label{sub:3.2}}

The condensate amplitudes $z_{i}$ are found by solving equation (\ref{eq12}),
where
the trapping potential is given by $\epsilon_{i}=\Omega (i-(I+1)/2)^{2}$ for
site $i$, with $\Omega=\frac{1}{2}m\omega^{2}a^{2}$, and $a$ is
the inter-lattice spacing. The quasi-particle amplitudes for site
$i$, $u_{i}^{q}$and $v_{i}^{q}$are found by solving equations (\ref{eq13}).

This set of equations can again be solved iteratively. In performing
the calculation, we actually 
set the $c^{q}$'s to zero when solving the Bogoliubov-de
Gennes equations, but do so in the Hartree-Fock basis, thus ensuring orthogonality
of the ground state and the excited states \cite{key-18}. The Hartree-Fock basis
is given by the normalised solutions to the eigenvalue problem given
by equation (\ref{eq12}), but where the ground state (zero energy
solution) is excluded. 

First, let us rewrite the Bogoliubov-de Gennes equations (\ref{eq13})
in matrix form\begin{equation}
\hbar\omega_{q}\left[\begin{array}{c}
\mathbf{u}^{q}\\
\mathbf{v}^{q}\end{array}\right]=\left[\begin{array}{cc}
\hat{\mathcal{L}} & M\\
-M & -\hat{\mathcal{L}}\end{array}\right]\left[\begin{array}{c}
\mathbf{u}^{q}\\
\mathbf{v}^{q}\end{array}\right],\label{eq61}\end{equation}
where\begin{equation}
\begin{array}{ccc}
\mathbf{u}^{q}=\left[\begin{array}{c}
u_{1}^{q}\\
\vdots\\
u_{i}^{q}\\
\vdots\\
u_{I}^{q}\end{array}\right] & \mathrm{,} & \mathbf{v}^{q}=\left[\begin{array}{c}
v_{1}^{q}\\
\vdots\\
v_{i}^{q}\\
\vdots\\
v_{I}^{q}\end{array}\right],\end{array}\label{eq60}\end{equation}
\begin{equation}
\hat{\mathcal{L}}=2V_{\rm{eff}}\left(n_{c_{i}}+\tilde{n}_{i}\right)-\mu+\Omega (i-(I+1)/2)^{2}-\hat{\mathcal{J}}\label{eq57}\end{equation}
and\begin{equation}
M=-V_{\rm{eff}}z_{i}^{2}\label{eq57a}\end{equation}
Here $\hat{\mathcal{J}}$ is defined as the operator acting on $z_{i}$, $u_{i}^{q}$,
and $v_{i}^{q}$ as follows:\begin{equation}
\hat{\mathcal{J}}u_{i}^{q}=u_{i+1}^{q}+u_{i-1}^{q}.\label{eq58}\end{equation}

Now, let $\left\{ z_{i}^{q}\right\} $constitute the eigenstates of
equation (\ref{eq12}) with eigenvalues $\mu^{q}$. We order the
eigenvalues into ascending order, and order the normalised eigenstates accordingly,
call these $\left\{ \xi_{i}^{q}\right\} $. The state $\xi_{i}^{0}$
corresponds to the Goldstone mode, we must exclude this in order to
obtain the Hartree-Fock-Bogoliubov basis. Let us define the matrix\begin{equation}
\mathbf{U}=\left[\begin{array}{ccccc}
\xi^{1} & \ldots & \xi^{q} & \ldots & \xi^{I-1}\end{array}\right],\label{eq55}\end{equation}
where\begin{equation}
\xi^{q}=\left[\begin{array}{c}
\xi_{1}^{q}\\
\vdots\\
\xi_{i}^{q}\\
\vdots\\
\xi_{I}^{q}\end{array}\right].\label{eq56}\end{equation}
 Let $\left\{ \mathbf{u}_{HFB}^{q},\mathbf{v}_{HFB}^{q}\right\} $be
the solution to the Bogoliubov-de Gennes equations in the Hartree-Fock-Bogoliubov
basis, then the solution to the matrix equation\begin{equation}
\hbar\omega_{q}\left[\begin{array}{c}
\mathbf{u}_{HFB}^{q}\\
\mathbf{v}_{HFB}^{q}\end{array}\right]=\left[\begin{array}{cc}
U^{T}\hat{\mathcal{L}}U & U^{T}MU\\
-U^{T}MU & -U^{T}\hat{\mathcal{L}}U\end{array}\right]\left[\begin{array}{c}
\mathbf{u}_{HFB}^{q}\\
\mathbf{v}_{HFB}^{q}\end{array}\right]\label{eq62}\end{equation}
gives the Bogoliubov quasiparticle amplitudes in the Hartree-Fock-Bogoliubov
basis. To obtain the Bogoliubov quasiparticle amplitudes$\left\{ u_{i}^{q},v_{i}^{q}\right\} $
one transforms back using\begin{equation}
\begin{array}{c}
\mathbf{u}^{q}=U\mathbf{u}_{HFB}^{q}\\
\mathbf{v}^{q}=U\mathbf{v}_{HFB}^{q}\end{array}.\label{eq63}\end{equation}

The iterative solution of these equations is then obtained numerically.

\section{Results}

We present results for a lattice of depth of approximately $16.8E_R$, where the recoil energy is defined as $E_R=h^{2}/8ma^{2}$ with $a$ the inter-site spacing. This is defined for the optical lattice parameters defined in reference \cite{key-20} and using the band structure calculations of \cite {key-21}.

\subsection{Translationally Invariant Lattice}

\begin{figure}
\includegraphics[%
  scale=0.70,
  angle=270]{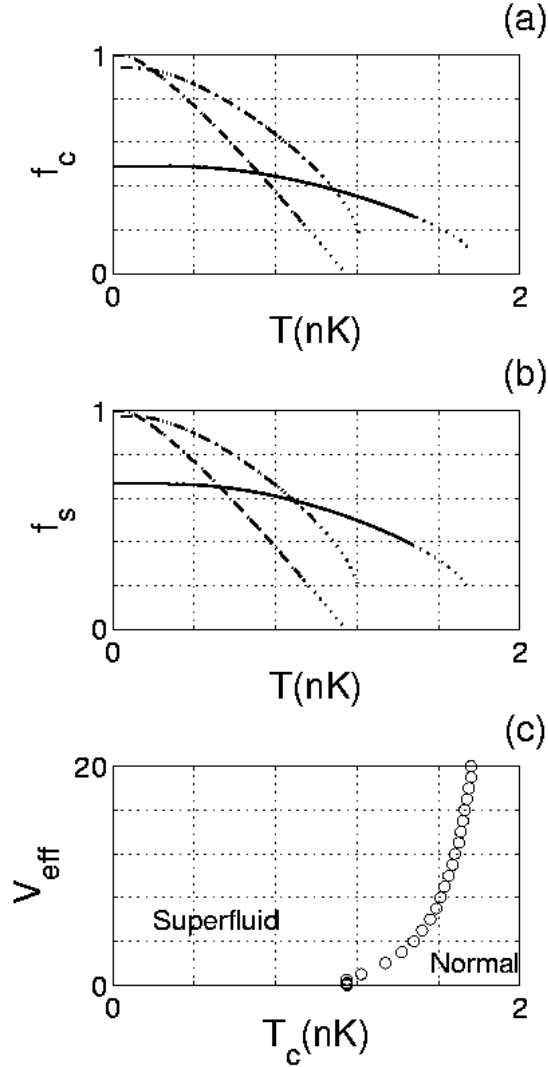}

\caption{Translationally invariant lattice with ten atoms and ten sites. Overall condensate (a) and superfluid (b) fractions as a function of temperature. The corresponding phase diagram 
is shown in panel (c).
\label{cap:fig1}}
\end{figure}

\begin{figure}
\includegraphics[%
  scale=0.7,
  angle=270]{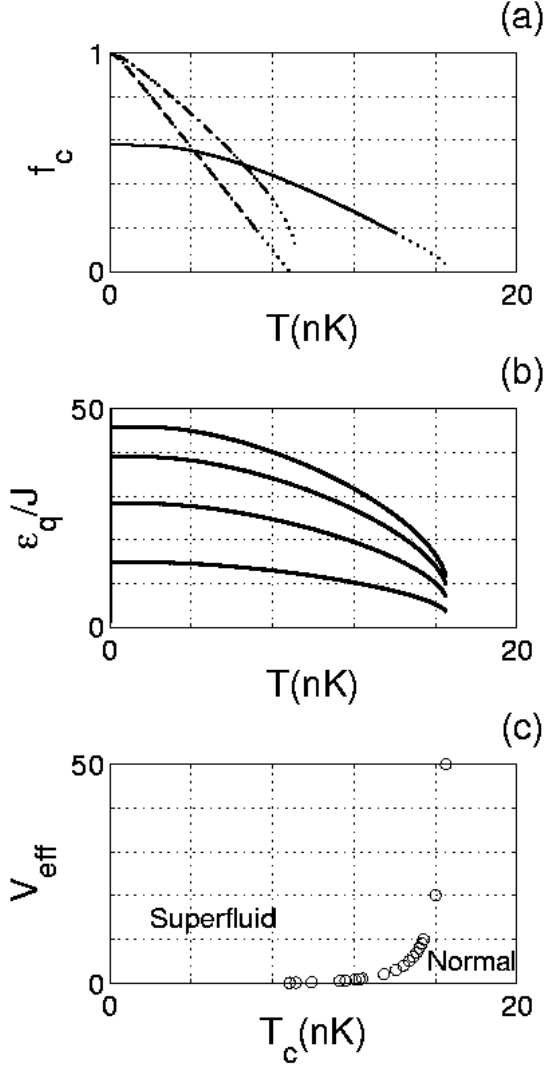}

\caption{Translationally invariant lattice with ten atoms per site. Condensate Fraction
(a), and
excitation spectrum as a function of temperature (b) with the corresponding phase diagram (c).
\label{cap:fig2}}
\end{figure}

\begin{figure}
\includegraphics[%
  scale=0.7,
  angle=270]{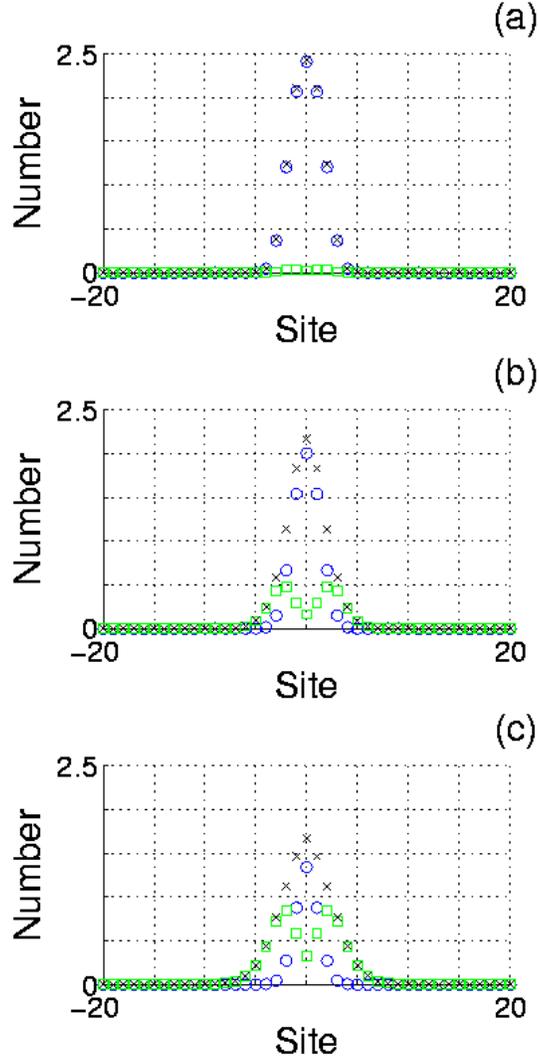}

\caption{Inhomogeneous lattice with forty one lattice sites and ten atoms in total for
$V_{\rm{eff}}=1$, at (a) $T=0$ nK, (b) $T=1$ nK (c)
$T=1.6$ nK. Circles represent the number of condensate atoms, squares the number of excited atoms and crosses the total number of atoms\label{cap:fig3}}
\end{figure}

\begin{figure}
\includegraphics[%
  scale=0.7,
  angle=270]{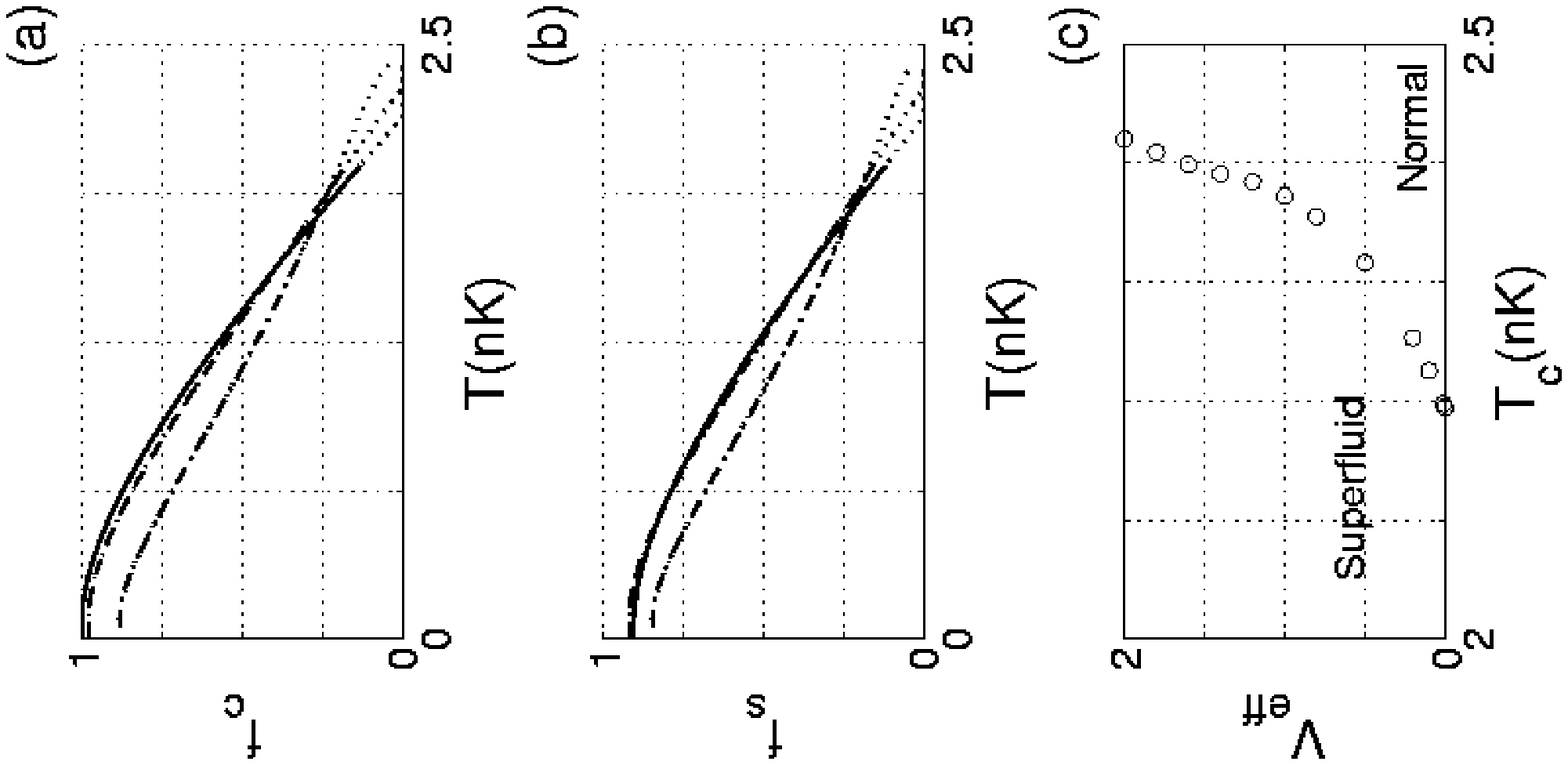}

\caption{Overall condensate and superfluid fractions as a
function of temperature, and the corresponding phase diagram for
an optical lattice in a harmonic potential with forty lattice sites (even case) and 
ten atoms. (a) corresponds to the condensate
fraction , (b) to the superfluid
fraction and (c) to the phase diagram.\label{cap:fig4}}
\end{figure}

In figures \ref{cap:fig1} and \ref{cap:fig2} we present results for a 
translationally invariant lattice with one atom and ten atoms per site
respectively for various values
of the effective interaction potential $V_{\rm{eff}}$. In both cases ten lattice sites, with
periodic boundary conditions, were used. One observes, a decrease in both the condensate
and superfluid fractions with temperature and we interpret the point at which these
densities approach zero as indicative of the critical temperature 
$T_{c}$ for the superfluid to normal gas phase transition. This interpretation
is supported by an examination of the low lying excitation spectrum which, for
the case $V_{\rm{eff}}=20$, is shown in panel (b) of figure \ref{cap:fig2}. The
``softening'' of the modes, indicative of the phase transition, is clearly seen and
coincides with
the transition temperature obtained by noting the temperature at which the condensate and
superfluid densities approach zero.

We obtain such a transition temperature for each value of the effective
interaction potential $V_{\rm{eff}}$, producing a phase diagram as shown in figures
\ref{cap:fig1}(c) and \ref{cap:fig2}(c), where the superfluid phase and the normal phase
are as indicated, the superfluid lying to the left of the curve. The
transition to the Mott insulator phase cannot be determined by this
analysis, but one would expect the transition from the superfluid
phase to the Mott insulator phase to occur when the site coupling
strength $J$ is decreased below a certain point, depending on the on-site interaction strength $V_{\rm{eff}}$. In practice the Mott insulator phase transition will occur when $V_{\rm{eff}}=V/J$ exceeds some critical value. Thus a transition to the Mott insulator phase is also possible by increasing the on-site interaction strength $V$ for a given coupling strength $J$.
We note here that, in all instances, the transition temperature increases
with the effective interaction potential, as is indeed the case with
a homogeneous Bose gas (ie. a Bose gas in the absence of an optical
lattice and of a confining potential)\cite{key-22,key-23}. This reentrant behaviour has also been
predicted for the translationally invariant lattice by Kleinert {\em et al.}
\cite{key-24}. Our results are consistent with their conclusions. In addition, they predict a
reduction of the critical temperature as the effective interaction strength is increased
further. We, however, are unable to explore this regime as it extends beyond the validity
of our model.

%It is interesting to note in figure \ref{cap:fig2}(c) that for $V_{\rm{eff}}\ll k_{B}T_{c}/3J$,
%the increase in transition temperature with increasing $V_{\rm{eff}}$ is
%very slight (virtually constant), but then for $V_{\rm{eff}}\sim k_{B}T_{c}/3J$,
%we notice a rapid increase in transition temperature with $V_{\rm{eff}}$
%and this continues to $V_{\rm{eff}}\sim k_{B}T_{c}/J$ whereupon the
%increase of $T_{c}$ with $V_{\rm{eff}}$ progressively diminishes with
%increasing $T_{c}$. 
%For $V_{\rm{eff}}\ll k_{B}T/3J$, the interactions are negligible,
%and we have effectively an ideal gas and the superfluid to normal phase transition 
%temperature is
%very nearly independent of $V_{\rm{eff}}$. We are in the very special regime of Bose-Einstein
%condensation phase transition in the absence of interactions\cite{footnote-1}. It is only when $V_{\rm{eff}}\sim %k_{B}T_{c}/3J$,
%that interactions start to become important and the phase transition is modified,
%yielding the rapid increase in the
%transition temperature $T_{c}$ as a function of the effective interaction
%potential $V_{\rm{eff}}$ in this regime. 

\subsection{Inhomogeneous Lattice}

In this section we present results for the case of an optical lattice in
the presence of a harmonic trapping potential. The condensate fraction
$f_{c}$ and the superfluid fraction $f_{s}$ are evaluated as a function
of the effective potential $V_{\rm{eff}}$ and the temperature. 

Figure \ref{cap:fig3} shows plots of the number
of condensate atoms, excited atoms and of the total number of atoms
for each lattice site for the case of an inhomogeneous optical lattice
consisting of forty one lattice sites (odd case) with ten atoms in
total, for $V_{\rm{eff}}=1$ at various temperatures ranging from $T=0$ nK
to $T=1.6$ nK. At zero temperature, the condensate atom distribution
is bell-shaped, peaked at the central lattice site.
There is a small quantum depletion even at zero temperature, and the
distribution of excited atoms is shaped as a bimodal distribution,
centred about the central lattice site. As the temperature increases, 
the condensate population decreases,
but the distribution still remains bell-shaped, and the excited population
increases.  In figure \ref{cap:fig4} we present the overall condensate
and superfluid fractions and the corresponding phase diagram
for optical lattices in a harmonic potential consisting of forty lattice sites 
(even case). Panel (a) corresponds to the condensate fraction, (b) the superfluid fraction, 
and (c) the phase diagram. We note that for higher temperatures, calculations performed using forty one lattice sites begin to show a marked difference. This is indicative of the
fact that we are pushing the bounds of validity 
of our model. In particular we are seeing significant finite size effects, which affect
the value of the chemical potential and, ultimately, a failure of the mean field
approximation. The dotted continuation of the lines in figure \ref{cap:fig4} indicate
where our calculations become unreliable, but are included for completness. We are still
able to use our model to obtain an estimate of the critical temperature and hence the
trend in its dependence upon the effective interaction strength.
We therefore conclude from figure \ref{cap:fig4} that the superfluid to
normal phase transition temperature increases with increasing $V_{\rm{eff}}$. 
It is clear, then, that
the shift in critical temperature with effective interaction potential
is positive definite for a Bose gas in a one-dimensional optical lattice,
regardless of whether the system is confined in a (harmonic) trapping
potential or not. This is in contrast to the case of a three-dimensional
Bose gas, where $\Delta T_{c}$ changes sign for the trapped gas.

\section{Conclusions}

We have applied the descretized Hartree-Fock-Bogoliubov formulation to the 
Bose-Hubbard model in order to calculate the dependence of the condensate and superfluid
fractions on the temperature. We have used this to estimate the critical temperature for
the superfluid to normal phase transition for both a translationally
invariant optical lattice (no external trap present), and an inhomogeneous
optical lattice (contained within an external harmonic trap).
This has enabled us to investigate the phase diagram for both cases
and we observe that the transition temperature increases with increasing
effective interaction potential $V_{\rm{eff}}$. Unlike the
homogeneous case with no optical lattice, this positive shift in the
critical temperature with interaction strength is present in both
the translationally invariant case and when a parabolic confining
potential is imposed. In the homogeneos gas the shift in the critical
temperature is only positive in
the absence of a confining potential. These conclusions are consistent with previous
work for the translationally invariant case\cite{key-24}, extending this result to
include parabolic confinement.

\section{Acknowledgements}
We would like to thank the Marsden Fund of the Royal Society of New
Zealand and the University of Otago for financial support.

\end{document}